\documentclass[preprint,showpacs,preprintnumbers,amsmath,amssymb]{revtex4}
\usepackage{amssymb}
\usepackage{graphicx}
\usepackage{dcolumn}
\usepackage{bm}

\makeatletter

\begin{document}

\title{Quantum Oscillations of Tunnel Magnetoresistance Induced by Spin-Wave
Excitations in Ferromagnet-Ferromagnet-Ferromagnet Double Barrier Tunnel
Junctions}
\author{Xi Chen}
\author{Qing-Rong Zheng}
\author{Gang Su}
\email[Author to whom correspondence should be addressed. ]{ Email:
gsu@gucas.ac.cn}
\affiliation{{College of Physical Sciences, Graduate University of Chinese Academy of
Sciences, P.O. Box 4588, Beijing 100049, China}}

\begin{abstract}
The possibility of quantum oscillations of the tunnel conductance and
magnetoresistance induced by spin-wave excitations in a
ferromagnet-ferromagnet-ferromagnet double barrier tunnel junction, when the
magnetizations of the two side ferromagnets are aligned antiparallel to that
of the middle ferromagnet, is investigated in a self-consistent manner by
means of Keldysh nonequilibrium Green function method. It has been found
that owing to the s-d exchange interactions between conduction electrons and
the spin density induced by spin accumulation in the middle ferromagnet, the
differential conductance and the TMR indeed oscillate with the increase of
bias voltage, being consistent with the phenomenon that is observed recently
in experiments. The effects of magnon modes, the energy levels of electrons
as well as the molecular field in the central ferromagnet on the oscillatory
transport property of the system are also discussed.
\end{abstract}

\maketitle

\section{Introduction}

In past decades, the spin-dependent transport properties in magnetic tunnel
junctions (MTJs) have been extensively investigated both experimentally and
theoretically, where a great progress has been made (see, e.g. Refs. \cite%
{Prinz,Wolf,Zutic,tserk,Su} for reviews). It has been unveiled that owing to
the conduction electron scatterings, the tunnel current through the MTJ is
modulated by the relative orientation of magnetizations, giving rise to the
so-called tunnel magnetoresistance (TMR) effect. As the quality of tunnel
junctions is being improved, a large TMR, which is expected by practical
applications, has been achieved in several systems. On the other hand, a
reverse effect of TMR, coined as the spin transfer effect \cite{Berger,J.C},
has also been proposed, which predicts that the orientation of magnetization
of free ferromagnetic layer can be switched by passing a spin-polarized
electrical current, and spin waves could also be excited. This latter effect
has been confirmed experimentally in a number of systems.

Although single barrier MTJs already show abundant characteristics
concerning the spin-dependent electrical transport, a double barrier
magnetic tunnel junction (DBMTJ), in which the formation of quantum well
states and the resonant tunneling phenomenon are theoretically anticipated,
has also attracted much attention in recent years \cite%
{Zhang,F.M,Zhang2,Sheng,Stein,zhu1,jin1,zhu2,Colis,Han,jin2,zhu3,mu1,jin3,jin4,mu2,Xing,Nozaki,Yan,Zeng}%
. In order to observe the coherent tunneling thru the DBMTJ, people have
attempted to improve the junction quality to eliminate the influences from
the interface roughness and impurity scattering, and remarkable advances
have been achieved on this aspect.

Recently, an unusual magnetotransport phenomenon in the
ferromagnet-ferromagnet-ferromagnet (FM-FM-FM) DBMTJs was reported by Zeng
\textit{et al}. \cite{Zeng}. They observed that, when the magnetization of
center (free) magnetic layer was antiparallel (AP) to the magnetization of
the two outer (pinned) magnetic layers, the conductance and TMR oscillate
distinctly with the applied bias voltage, while for the parallel (P)
situation, no such oscillation was seen. Unlike the previous oscillatory
tunnel magnetoresistance, this unusual phenomenon can neither be explained
by Coulomb blockade effect since the middle FM layer is continuous, the
charge effect should be equal in P and AP configurations and the charging
energy is negligibly small, nor be attributed to the resonant tunneling,
because the observed period of oscillation is too small to account for the
energy level spacing of the quantum well states. Considering that the
conductance oscillation is asymmetrical for P and AP configurations, and the
energy level of the unusual phenomenon is the same as the typical energy of
a magnon, one may speculate that the unusual oscillation behavior could be
induced by the magnon-assisted tunneling \cite{Zeng}. This is because in the
AP state, the nonequilibrium spin density, which is proportional to the
applied bias, could be accumulated near the interfaces in the middle region
to emit spin waves, and the magnon-assisted tunneling would contribute to
the conductance, while in the P state the spin wave emission is forbidden
due to the spin angular momentum conservation, as discussed previously \cite%
{S.Zhang}.

As there is no previous theoretical study devoting to the investigation on
the possible quantum oscillations induced by spin wave excitations, in this
paper, by using the nonequilibrium Green function method, we shall examine
theoretically the above-mentioned idea by studying the possibility of
magnon-assisted tunneling in the FM-FM-FM DBMTJ, and explore whether the
magnon-assisted tunneling could really cause the oscillations of the
differential conductance and TMR with the applied bias voltage.

The rest of this paper is organized as follows: In Sec. II, a model is
proposed. The tunnel current and relevant Green functions are obtained in
terms of the nonequilibrium Green function technique in Sec. III. In Sec.
IV, the transport properties of the system are numerically investigated, and
some discussions are presented. Finally, a brief summary is given in Sec. V.

\section{Model}

Let us consider a FM-FM-FM DBMTJ with three FM layers separated by two thin
insulating films. Suppose that the left (L) and right (R) FM electrodes with
magnetizations aligned parallel are applied by bias voltages $-V/2$ and $V/2$%
, respectively. The magnetization of the middle FM layer is presumed to be
antiparallel to those of the L and R electrodes so that spin waves can be
emitted in the middle FM layer because of spin accumulation. The schematic
layout of this system is depicted in the inset of Fig. 1(a). The Hamiltonian
of the system reads

\begin{equation}
H=H_{L}+H_{R}+H_{C}+H_{LC}+H_{CR},  \label{hamil}
\end{equation}%
with
\begin{equation}
H_{\alpha }=\sum_{k_{\alpha }\sigma }\varepsilon _{k_{\alpha }\sigma
}a_{k_{\alpha }\sigma }^{+}a_{k_{\alpha }\sigma },\text{ }(\alpha =L,R)
\label{h-alpha}
\end{equation}

\begin{equation}
H_{C}=\underset{k\sigma }{\sum }\varepsilon _{k\sigma }c_{k\sigma
}^{+}c_{k\sigma }+\underset{q}{\sum }\hbar \omega _{q}b_{q}^{+}b_{q},
\label{h-c}
\end{equation}

\begin{eqnarray}
H_{LC} &=&\underset{k_{L}k\sigma }{\sum }T_{k_{L}k}^{d}(a_{k_{L}\sigma
}^{+}c_{k\sigma }+h.c.)  \notag \\
&&+\frac{1}{\sqrt{N}}\underset{k_{L}kq}{\sum }T_{k_{L}kq}^{J}S(q)(a_{k_{L}%
\uparrow }^{+}c_{k\uparrow }-a_{k_{L}\downarrow }^{+}c_{k\downarrow
}+c_{k\uparrow }^{+}a_{k_{L}\uparrow }-c_{k\downarrow
}^{+}a_{k_{L}\downarrow })  \notag \\
&&+\frac{1}{\sqrt{N}}\underset{k_{L}kq}{\sum }T_{k_{L}kq}^{J}\sqrt{2S}%
(a_{k_{L}\uparrow }^{+}c_{k\downarrow }b_{q}^{+}+c_{k\uparrow
}^{+}a_{k_{L}\downarrow }b_{q}^{+}+a_{k_{L}\downarrow }^{+}c_{k\uparrow
}b_{q}+c_{k\downarrow }^{+}a_{k_{L}\uparrow }b_{q}),  \label{h-LC}
\end{eqnarray}

\begin{eqnarray}
H_{CR} &=&\underset{k_{R}k\sigma }{\sum }T_{k_{R}k}^{d}(a_{k_{R}\sigma
}^{+}c_{k\sigma }+h.c.)  \notag \\
&&+\frac{1}{\sqrt{N}}\underset{k_{R}kq}{\sum }T_{k_{R}kq}^{J}S(q)(a_{k_{R}%
\uparrow }^{+}c_{k\uparrow }-a_{k_{R}\downarrow }^{+}c_{k\downarrow
}+c_{k\uparrow }^{+}a_{k_{R}\uparrow }-c_{k\downarrow
}^{+}a_{k_{R}\downarrow })  \notag \\
&&+\frac{1}{\sqrt{N}}\underset{k_{R}kq}{\sum }T_{k_{R}kq}^{J}\sqrt{2S}%
(a_{k_{R}\uparrow }^{+}c_{k\downarrow }b_{q}^{+}+c_{k\uparrow
}^{+}a_{k_{R}\downarrow }b_{q}^{+}+a_{k_{R}\downarrow }^{+}c_{k\uparrow
}b_{q}+c_{k\downarrow }^{+}a_{k_{R}\uparrow }b_{q}),  \label{h-CR}
\end{eqnarray}%
where $a_{k_{\alpha }\sigma }$ and $c_{k\sigma }$ are annihilation operators
of electrons with momentum $k$ and spin $\sigma $ in the $\alpha $ electrode
and in the middle FM layer, respectively, $\varepsilon _{k_{\alpha }\sigma
}=\varepsilon _{k_{\alpha }}-\sigma {}M_{\alpha }-eV_{\alpha }$ with $%
\varepsilon _{k_{\alpha }}$ the single-electron energy and $M_{\alpha }$ the
molecular field in the $\alpha $ electrode, $\varepsilon _{k\sigma
}=\varepsilon _{k}-\sigma M$ with $\varepsilon _{k}$ the single-electron
energy and $M$ the molecular field in the middle FM layer, $b_{q}$ is the
annihilation operator of magnon with momentum $q$ in the middle region, $%
\hbar \omega _{q}$ is the magnon energy, $N=\underset{q}{\sum }\langle
n_{q}^{s}\rangle $ with $n_{q}^{s}=b_{q}^{+}b_{q}$ is the number of magnons,
$S(q)=S-n_{q}^{s}$ where $S=1/2$ is the spin of electron, ${T_{k_{\alpha
}k}^{d}}$ are tunneling matrix elements of electrons between the $\alpha $
electrode and middle FM layer, $T_{k_{\alpha }kq}^{J}$ are coupling matrix
elements between the electrons in $\alpha $ electrode and magnons in the
middle FM region.

It is noting that $H_{LC}$ ($H_{CR}$) describes the coupling between
electrons in the L (R) electrode and electrons as well as magnons in the
central FM region, where the terms containing $T_{k_{\alpha }kq}^{J}$ in
Eqs. (\ref{h-LC}) and (\ref{h-CR}) are due to the \textit{s-d} exchange
interactions\cite{S.Zhang}. Without loss of generality, we further assume $%
T_{k_{L}kq}^{J}=\gamma {T_{k_{L}k}^{d}}$ in the following discussions.

\section{Tunnel Current and Green Functions}

\subsection{Tunnel Current}

Starting from Eq. (\ref{hamil}), after some cubersome but straightforward
calculations, one may obtain the tunnel electrical current $I$

\begin{equation}
I=I_{L\uparrow }+I_{L_{\downarrow }},  \label{current}
\end{equation}

\begin{eqnarray*}
I_{L\uparrow }(t) &=&\frac{2e}{\hbar }\Re
e[\sum_{k_{L}k}T_{k_{L}k}^{d}G_{k\uparrow k_{L}\uparrow }^{<}(t,t)+\frac{1}{%
\sqrt{N}}\underset{k_{L}kq}{\sum }T_{k_{L}kq}^{J}(S-\langle n_{q}^{s}\rangle
)G_{k\uparrow k_{L}\uparrow }^{<}(t,t) \\
&&+\sqrt{\frac{2S}{N}}\underset{k_{L}kq}{\sum }T_{k_{L}kq}^{J}G_{k\downarrow
k_{L}\uparrow }^{q<}(t,t)],
\end{eqnarray*}

\begin{eqnarray*}
I_{L_{\downarrow }}(t) &=&\frac{2e}{\hbar }\Re
e[\sum_{k_{L}k}T_{k_{L}k}^{d}G_{k\downarrow k_{L}\downarrow }^{<}(t,t)-\frac{%
1}{\sqrt{N}}\underset{k_{L}kq}{\sum }T_{k_{L}kq}^{J}(S-\langle
n_{q}^{s}\rangle )G_{k\downarrow k_{L}\downarrow }^{<}(t,t) \\
&&+\sqrt{\frac{2S}{N}}\underset{k_{L}kq}{\sum }T_{k_{L}kq}^{J}G_{k\uparrow
k_{L}\downarrow }^{q<}(t,t)],
\end{eqnarray*}
where the lesser Green functions are defined as%
\begin{equation}
G_{k\sigma k_{L}\sigma }^{<}(t,t^{\prime })=i\langle a_{k_{L}\sigma
}^{+}(t^{\prime })c_{k\sigma }(t)\rangle ,  \label{G-L}
\end{equation}

\begin{equation}
G_{k\downarrow k_{L}\uparrow }^{q<}(t,t^{\prime })=i\langle a_{k_{L}\uparrow
}^{+}(t^{\prime })b_{q}^{+}(t^{\prime })c_{k\downarrow }(t)\rangle ,
\label{G-qL-up}
\end{equation}

\begin{equation}
G_{k\uparrow k_{L}\downarrow }^{q<}(t,t^{\prime })=i\langle
a_{k_{L}\downarrow }^{+}(t^{\prime })b_{q}(t^{\prime })c_{k\uparrow
}(t)\rangle .
\end{equation}%
It should be remarked that in the above derivations, we have made decoupling
approximations for the terms containing $n_{q}^{s}$ to simplify the
calculations. From these above equations, one may see that to get the tunnel
electrical current, the lesser Green functions must be obtained. In the
following, we shall employ Keldysh's nonequilibrium Green function method to
get all self-consistent equations to determine the lesser Green functions.
As the lesser Green function is closely related to the retarded and advanced
Green functions according to Keldysh formalism, the relevant retarded and
advanced Green functions of electrons and magnons should be first calculated.

Accordingly, the differential tunnel conductance ($G$) is obtained by $%
G(V)=dI(V)/dV$, and the TMR can be calculated by $TMR=(1-G_{\uparrow
\downarrow }/G_{\uparrow \uparrow })$ where $G_{\uparrow \downarrow }$ ($%
G_{\uparrow \uparrow }$) is the differential conductance when the
magnetizations of the middle FM and the side FM are aligned antiparallel
(parallel).

\subsection{Green Functions of Electrons}

Let us define useful retarded Green functions for electrons as

\begin{equation}
G_{k\sigma k^{\prime }\sigma }^{r}(t,t^{\prime })=-i\theta (t-t^{\prime
})\langle \{c_{k\sigma }(t),c_{k^{\prime }\sigma }^{+}(t^{\prime })\}\rangle
,  \label{G-r}
\end{equation}

\begin{equation}
G_{k\sigma k_{\alpha }\sigma }^{r}(t,t^{\prime })=-i\theta (t-t^{\prime
})\langle \{c_{k\sigma }(t),a_{k_{\alpha }\sigma }^{+}(t^{\prime })\}\rangle
,  \label{G-r-a}
\end{equation}

\begin{equation}
G_{k\uparrow k_{\alpha }\downarrow }^{r(q)}(t,t^{\prime })=-i\theta
(t-t^{\prime })\langle \{c_{k\uparrow }(t),a_{k_{\alpha }\downarrow
}^{+}(t^{\prime })b_{q}(t^{\prime })\}\rangle ,  \label{G-r-ab}
\end{equation}

\begin{equation}
G_{k\uparrow k^{\prime }\downarrow }^{r(q)}(t,t^{\prime })=-i\theta
(t-t^{\prime })\langle \{c_{k\uparrow }(t),c_{k^{\prime }\downarrow
}^{+}(t^{\prime })b_{q}(t^{\prime })\}\rangle ,  \label{G-r-cb}
\end{equation}

\begin{equation}
G_{k\downarrow k_{\alpha }\uparrow }^{r(q)}(t,t^{\prime })=-i\theta
(t-t^{\prime })\langle \{c_{k\downarrow }(t),a_{k_{\alpha }\uparrow
}^{+}(t^{\prime })b_{q}^{+}(t^{\prime })\}\rangle ,  \label{G-r-ab+}
\end{equation}

\begin{equation}
G_{k\downarrow k^{\prime }\uparrow }^{r(q)}(t,t^{\prime })=-i\theta
(t-t^{\prime })\langle \{c_{k\downarrow }(t),c_{k^{\prime }\uparrow
}^{+}(t^{\prime })b_{q}^{+}(t^{\prime })\}\rangle .  \label{G-r-cb+}
\end{equation}

In terms of the equation of motion, after a tedious calculation, up to the
third-order of Green functions, we get the following equations

\begin{eqnarray*}
(\varepsilon -\varepsilon _{k_{\alpha }\uparrow })G_{k\uparrow k_{\alpha
}\uparrow }^{r}(\varepsilon ) &=&[T^{d}+\frac{T^{J}}{\sqrt{N}}{\sum_{q}}%
(S-\langle n_{q}^{s}\rangle )]{\sum_{k^{\prime }}}G_{k\uparrow k^{\prime
}\uparrow }^{r}(\varepsilon ) \\
&&+\sqrt{\frac{2S}{N}}T^{J}\underset{k^{\prime }q}{\sum }G_{k\uparrow
k^{\prime }\downarrow }^{r(q)}(\varepsilon ),
\end{eqnarray*}

\begin{eqnarray*}
(\varepsilon -\varepsilon _{k^{\prime }\uparrow })G_{k\uparrow k^{\prime
}\uparrow }^{r}(\varepsilon ) &=&[T^{d}+\frac{T^{J}}{\sqrt{N}}{\sum_{q}}%
(S-\langle n_{q}^{s}\rangle )]\sum_{\alpha }\sum_{k_{\alpha }}G_{k\uparrow
k_{\alpha }\uparrow }^{r}(\varepsilon ) \\
&&+\sqrt{\frac{2S}{N}}T^{J}\underset{\alpha }{\sum }\underset{k_{\alpha }q}{%
\sum }G_{k\uparrow k_{\alpha }\downarrow }^{r(q)}(\varepsilon ),
\end{eqnarray*}

\begin{eqnarray*}
(\varepsilon -\varepsilon _{k_{\alpha }\downarrow }+\hbar \omega
_{q})G_{k\uparrow k_{\alpha }\downarrow }^{r(q)}(\varepsilon ) &=&[T^{d}-%
\frac{T^{J}}{\sqrt{N}}\sum_{q^{\prime }}(S-\langle n_{q^{\prime
}}^{s}\rangle )+\frac{T^{J}}{\sqrt{N}}\langle n_{k_{\alpha }\downarrow
}\rangle ]\sum_{k^{\prime }}G_{k\uparrow k^{\prime }\downarrow
}^{r(q)}(\varepsilon ) \\
&&+\sqrt{\frac{2S}{N}}T^{J}(\langle n_{q}^{s}\rangle +\langle n_{k_{\alpha
}\downarrow }\rangle )\sum_{k^{\prime }}G_{k\uparrow k^{\prime }\uparrow
}^{r}(\varepsilon ),
\end{eqnarray*}

\begin{eqnarray*}
(\varepsilon -\varepsilon _{k^{\prime }\downarrow }+\hbar \omega
_{q})G_{k\uparrow k^{\prime }\downarrow }^{r(q)}(\varepsilon ) &=&\left\{
T^{d}-\frac{T^{J}}{\sqrt{N}}[\sum_{q^{\prime }}(S-\langle n_{q^{\prime
}}^{s}\rangle )-\sum_{k^{\prime \prime }}\langle c_{k^{\prime }\downarrow
}^{+}c_{k^{\prime \prime }\downarrow }\rangle ]\right\} \sum_{\alpha
}\sum_{k_{\alpha }}G_{k\uparrow k_{\alpha }\downarrow }^{r(q)}(\varepsilon )
\\
&&+\sqrt{\frac{2S}{N}}T^{J}(\langle n_{q}^{s}\rangle +\sum_{k^{\prime \prime
}}\langle c_{k^{\prime }\downarrow }^{+}c_{k^{\prime \prime }\downarrow
}\rangle )\sum_{\alpha }\sum_{k_{\alpha }}G_{k\uparrow k_{\alpha }\uparrow
}^{r}(\varepsilon ),
\end{eqnarray*}%
\begin{eqnarray*}
(\varepsilon -\varepsilon _{k_{\alpha }\downarrow })G_{k\downarrow k_{\alpha
}\downarrow }^{r}(\varepsilon ) &=&[T^{d}-\frac{T^{J}}{\sqrt{N}}%
\sum_{q}(S-\langle n_{q}^{s}\rangle )]\sum_{k^{\prime }}G_{k\downarrow
k^{\prime }\downarrow }^{r}(\varepsilon ) \\
&&+\sqrt{\frac{2S}{N}}T^{J}\sum_{k^{\prime }q}G_{k\downarrow k^{\prime
}\uparrow }^{r(q)}(\varepsilon ),
\end{eqnarray*}

\begin{eqnarray*}
(\varepsilon -\varepsilon _{k^{\prime }\downarrow })G_{k\downarrow k^{\prime
}\downarrow }^{r}(\varepsilon ) &=&[T^{d}-\frac{T^{J}}{\sqrt{N}}%
\sum_{q}(S-\langle n_{q}^{s}\rangle )]\underset{\alpha }{\sum }\underset{%
k_{\alpha }}{\sum }G_{k\downarrow k_{\alpha }\downarrow }^{r}(\varepsilon )
\\
&&+\sqrt{\frac{2S}{N}}T^{J}\underset{\alpha }{\sum }\underset{k_{\alpha }q}{%
\sum }G_{k\downarrow k_{\alpha }\uparrow }^{r(q)}(\varepsilon ),
\end{eqnarray*}

\begin{eqnarray*}
(\varepsilon -\varepsilon _{k_{\alpha }\uparrow }-\hbar \omega
_{q})G_{k\downarrow k_{\alpha }\uparrow }^{r(q)}(\varepsilon ) &=&[T^{d}+%
\frac{T^{J}}{\sqrt{N}}\sum_{q^{\prime }}(S-\langle n_{q^{\prime
}}^{s}\rangle )+\frac{T^{J}}{\sqrt{N}}\langle n_{k_{\alpha }\uparrow
}\rangle ]\sum_{k^{\prime }}G_{k\downarrow k^{\prime }\uparrow
}^{r(q)}(\varepsilon ) \\
&&+\sqrt{\frac{2S}{N}}T^{J}(\langle n_{q}^{s}\rangle +1-\langle n_{k_{\alpha
}\uparrow }\rangle )\sum_{k^{\prime }}G_{k\downarrow k^{\prime }\downarrow
}^{r}(\varepsilon ),
\end{eqnarray*}

\begin{eqnarray*}
(\varepsilon -\varepsilon _{k^{\prime }\uparrow }-\hbar \omega
_{q})G_{k\downarrow k^{\prime }\uparrow }^{r}(\varepsilon ) &=&\left\{ T^{d}+%
\frac{T^{J}}{\sqrt{N}}[\sum_{q^{\prime }}(S-\langle n_{q^{\prime
}}^{s}\rangle )+\underset{k^{\prime \prime }}{\sum }\langle c_{k^{\prime
}\uparrow }^{+}c_{k^{\prime \prime }\uparrow }\rangle ]\right\} \underset{%
\alpha }{\sum }\underset{k_{\alpha }}{\sum }G_{k\downarrow k_{\alpha
}\uparrow }^{r(q)}(\varepsilon ) \\
&&+\sqrt{\frac{2S}{N}}T^{J}(\langle n_{q}^{s}\rangle +1-\underset{k^{\prime
\prime }}{\sum }\langle c_{k^{\prime }\uparrow }^{+}c_{k^{\prime \prime
}\uparrow }\rangle )\underset{\alpha }{\sum }\underset{k_{\alpha }}{\sum }%
G_{k\uparrow k_{\alpha }\uparrow }^{r}(\varepsilon ),
\end{eqnarray*}%
where we have presumed, for simplicity, the coupling matrix elements $%
T_{k_{L}k}^{d}$ and $T_{k_{L}kq}^{J}$ independent of momentum by considering
that only those electrons near the Fermi surface participate in the
transport process, and $n_{k_{\alpha }\sigma }=a_{k_{\alpha }\sigma
}^{+}a_{k_{\alpha }\sigma }$.

From these equations, the required Green functions can be obtained
self-consistently. On the other hand, the lesser self-energy $\Sigma ^{<}$
can be approximated by Ng's ansatz \cite{Ng}: $\Sigma ^{<}=\Sigma
_{0}^{<}(\Sigma _{0}^{r}-\Sigma _{0}^{a})^{-1}(\Sigma ^{r}-\Sigma ^{a})$,
where $\Sigma ^{r}-\Sigma ^{a}=G^{a-1}-G^{r-1}$, $\Sigma _{0}^{r}$ and $%
\Sigma _{0}^{<}$ are given by the following equations
\begin{equation}
\left(
\begin{array}{cc}
\Sigma _{0\uparrow \uparrow }^{r}(\varepsilon ) & \Sigma _{0\downarrow
\uparrow }^{r}(\varepsilon ) \\
\Sigma _{0\uparrow \downarrow }^{r}(\varepsilon ) & \Sigma _{0\downarrow
\downarrow }^{r}(\varepsilon )%
\end{array}%
\right) =\left(
\begin{array}{cc}
-\frac{i\Gamma _{L\uparrow }}{2}-\frac{i\Gamma _{R\uparrow }}{2} & 0 \\
0 & -\frac{i\Gamma _{L\downarrow }}{2}-\frac{i\Gamma _{R\downarrow }}{2}%
\end{array}%
\right) ,
\end{equation}

\begin{equation*}
\left(
\begin{array}{cc}
\Sigma _{0\uparrow \uparrow }^{<}(\varepsilon ) & \Sigma _{0\downarrow
\uparrow }^{<}(\varepsilon ) \\
\Sigma _{0\uparrow \downarrow }^{<}(\varepsilon ) & \Sigma _{0\downarrow
\downarrow }^{<}(\varepsilon )%
\end{array}%
\right) =\left(
\begin{array}{cc}
i\Gamma _{L\uparrow }f(\varepsilon -\frac{eV}{2})+i\Gamma _{R\uparrow
}f(\varepsilon +\frac{eV}{2}) & 0 \\
0 & i\Gamma _{L\downarrow }f(\varepsilon -\frac{eV}{2})+i\Gamma
_{R\downarrow }f(\varepsilon +\frac{eV}{2})%
\end{array}%
\right) ,
\end{equation*}%
where $\Gamma _{\alpha \sigma }(\varepsilon )$ is the linewidth function
defined by $\Gamma _{\alpha \sigma }(\varepsilon )=2\pi \underset{k_{\alpha }%
}{\sum }\rho _{\sigma }(k_{\alpha })\left\vert T_{k_{L}k}^{d}\right\vert
^{2} $ with $\rho _{\sigma }(k_{\alpha })$ the density of states of
electrons with momentum $k_{\alpha }$ and spin $\sigma $ in the $\alpha $th
FM electrode, and $f(\varepsilon )$ is the Fermi distribution function. By
means of $G^{<}=G^{r}\Sigma ^{<}G^{a}$, the lesser Green functions can be
procured.

\subsection{Green Functions of Magnons}

As the number of magnons, $N$, enters into the formalism, we need to obtain
the Green functions of magnons to determine $N$ self-consistently. Define
the retarded Green function of magnons as
\begin{equation}
G_{qq}^{r}(t,t^{^{\prime }})=-i\theta (t-t^{^{\prime }})\langle \lbrack
b_{q}(t),b_{q}^{+}(t^{^{\prime }})]\rangle .  \label{G-magnon}
\end{equation}%
By using the equation of motion, we have

\begin{eqnarray}
(\varepsilon -\hbar \omega _{q})G_{qq}^{r}(\varepsilon ) &=&1+\sqrt{\frac{2S%
}{N}}\sum\limits_{\alpha }\underset{k_{\alpha }k}{\sum }T_{k_{\alpha
}kq}^{J}[G_{kq}^{r(1)}(\varepsilon )+G_{kq}^{r(2)}(\varepsilon )]  \notag \\
&&+\frac{1}{\sqrt{N}}\sum\limits_{\alpha }\underset{k_{\alpha }k}{\sum }%
T_{k_{\alpha }kq}^{J}[-G_{kq}^{r(3)}(\varepsilon )+G_{kq}^{r(5)}(\varepsilon
)-G_{kq}^{r(4)}(\varepsilon )+G_{kq}^{r(6)}(\varepsilon )],
\label{G-m-motion}
\end{eqnarray}%
where $G_{kq}^{r(i)}(\varepsilon )$ ($i=1,\cdots ,5$) are the Fourier
transforms of the Green functions defined as below
\begin{eqnarray*}
G_{kq}^{r(1)}(t,t^{\prime }) &=&-i\theta (t-t^{\prime })\langle \lbrack
b_{q}(t),a_{k_{\alpha }\downarrow }^{+}(t^{\prime })c_{k\uparrow }(t^{\prime
})]\rangle , \\
G_{kq}^{r(2)}(t,t^{\prime }) &=&-i\theta (t-t^{\prime })\langle \lbrack
b_{q}(t),c_{k\downarrow }^{+}(t^{\prime })a_{k_{\alpha }\uparrow }(t^{\prime
})]\rangle , \\
G_{kq}^{r(3)}(t,t^{\prime }) &=&-i\theta (t-t^{\prime })\langle \lbrack
b_{q}(t),b_{q}^{+}(t^{^{\prime }})a_{k_{\alpha }\uparrow }^{+}(t^{^{\prime
}})c_{k\uparrow }(t^{^{\prime }})]\rangle , \\
G_{kq}^{r(4)}(t,t^{\prime }) &=&-i\theta (t-t^{\prime })\langle \lbrack
b_{q}(t),b_{q}^{+}(t^{^{\prime }})c_{k\uparrow }^{+}(t^{^{\prime
}})a_{k_{\alpha }\uparrow }(t^{^{\prime }})]\rangle , \\
G_{kq}^{r(5)}(t,t^{\prime }) &=&-i\theta (t-t^{\prime })\langle \lbrack
b_{q}(t),b_{q}^{+}(t^{^{\prime }})a_{k_{\alpha }\downarrow }^{+}(t^{^{\prime
}})c_{k\downarrow }(t^{^{\prime }})]\rangle , \\
G_{kq}^{r(6)}(t,t^{\prime }) &=&-i\theta (t-t^{\prime })\langle \lbrack
b_{q}(t),b_{q}^{+}(t^{^{\prime }})c_{k\downarrow }^{+}(t^{^{\prime
}})a_{k_{\alpha }\downarrow }(t^{^{\prime }})]\rangle .
\end{eqnarray*}%
By using repeatedly the equation of motion, and making appropriate cut-off
approximations, up to the second order, we have

\begin{equation}
G_{kq}^{r(1)}(\varepsilon )=-T^{J}\sqrt{\frac{2S}{N}}\frac{\langle
n_{k_{\alpha }\downarrow }\rangle -\sum_{k^{\prime }}\langle c_{k^{\prime
}\uparrow }^{+}c_{k\uparrow }\rangle }{\varepsilon -\varepsilon _{k_{\alpha
}\downarrow }+\varepsilon _{k\uparrow }}G_{qq}^{r}(\varepsilon ),
\label{G-r-1}
\end{equation}

\begin{equation}
G_{kq}^{r(2)}(\varepsilon )=T^{J}\sqrt{\frac{2S}{N}}\frac{\langle
n_{k_{\alpha }\uparrow }\rangle -\sum_{k^{\prime }}\langle c_{k\downarrow
}^{+}c_{k^{\prime }\downarrow }\rangle }{\varepsilon +\varepsilon
_{k_{\alpha }\uparrow }-\varepsilon _{k\downarrow }}G_{qq}^{r}(\varepsilon ),
\label{G-r-2}
\end{equation}

\begin{equation}
G_{kq}^{r(3)}(\varepsilon )=-\{\frac{\frac{T^{J}}{\sqrt{N}}(1-\langle
n_{k_{\alpha }\uparrow }\rangle )\sum_{k^{\prime }}\langle c_{k^{\prime
}\uparrow }^{+}c_{k\uparrow }\rangle -[T^{d}+\frac{T^{J}}{\sqrt{N}}%
\sum_{q^{\prime }}(S-\langle n_{q^{\prime }}^{s}\rangle )](\sum_{k^{\prime
}}\langle c_{k^{\prime }\uparrow }^{+}c_{k\uparrow }\rangle -\langle
n_{k_{\alpha }\uparrow }\rangle )}{\varepsilon -\hbar \omega
_{q}-\varepsilon _{k_{\alpha }\uparrow }+\varepsilon _{k\uparrow }}%
\}G_{qq}^{r}(\varepsilon ),  \label{G-r-3}
\end{equation}

\begin{equation}
G_{kq}^{r(4)}(\varepsilon )=-\{\frac{\frac{T^{J}}{\sqrt{N}}%
(1-\sum_{k^{\prime }}\langle c_{k\uparrow }^{+}c_{k^{\prime }\uparrow
}\rangle )\langle n_{k_{\alpha }\uparrow }\rangle +[T^{d}+\frac{T^{J}}{\sqrt{%
N}}\sum_{q^{\prime }}(S-\langle n_{q^{\prime }}^{s}\rangle
)](\sum_{k^{\prime }}\langle c_{k\uparrow }^{+}c_{k^{\prime }\uparrow
}\rangle -\langle n_{k_{\alpha }\uparrow }\rangle )}{\varepsilon -\hbar
\omega _{q}+\varepsilon _{k_{\alpha }\uparrow }-\varepsilon _{k\uparrow }}%
\}G_{qq}^{r}(\varepsilon ),  \label{G-r-4}
\end{equation}

\begin{equation}
G_{kq}^{r(5)}(\varepsilon )=\{\frac{\frac{T^{J}}{\sqrt{N}}(1-\langle
n_{k_{\alpha }\downarrow }\rangle )\sum_{k^{\prime }}\langle c_{k^{\prime
}\downarrow }^{+}c_{k\downarrow }\rangle +[T^{d}-\frac{T^{J}}{\sqrt{N}}%
\sum_{q^{\prime }}(S-\langle n_{q^{\prime }}^{s}\rangle )](\sum_{k^{\prime
}}\langle c_{k^{\prime }\downarrow }^{+}c_{k\downarrow }\rangle -\langle
n_{k_{\alpha }\downarrow }\rangle )}{\varepsilon -\hbar \omega
_{q}-\varepsilon _{k_{\alpha }\downarrow }+\varepsilon _{k\downarrow }}%
\}G_{qq}^{r}(\varepsilon ),  \label{G-r-5}
\end{equation}

\begin{equation}
G_{kq}^{r(6)}(\varepsilon )=\{\frac{\frac{T^{J}}{\sqrt{N}}(1-\sum_{k^{\prime
}}\langle c_{k\downarrow }^{+}c_{k^{\prime }\downarrow }\rangle )\langle
n_{k_{\alpha }\downarrow }\rangle -[T^{d}-\frac{T^{J}}{\sqrt{N}}%
\sum_{q^{\prime }}(S-\langle n_{q^{\prime }}^{s}\rangle )](\sum_{k^{\prime
}}\langle c_{k\downarrow }^{+}c_{k^{\prime }\downarrow }\rangle -\langle
n_{k_{\alpha }\downarrow }\rangle )}{\varepsilon -\hbar \omega
_{q}+\varepsilon _{k_{\alpha }\downarrow }-\varepsilon _{k\downarrow }}%
\}G_{qq}^{r}(\varepsilon ).  \label{G-r-6}
\end{equation}%
The number of magnons can thus be obtained by the spectral theorem
\begin{equation}
N=\sum\limits_{q}\langle n_{q}^{s}\rangle =\sum\limits_{q}\Im m\int \frac{%
d\varepsilon }{2\pi }G_{qq}^{<}(\varepsilon )=\sum\limits_{q}\int \frac{%
d\varepsilon }{2\pi }f_{s}(\varepsilon )[G_{qq}^{r}(\varepsilon
)-G_{qq}^{a}(\varepsilon )],  \label{N-magnon}
\end{equation}%
where $f_{s}(\varepsilon )$ is the Bose distribution function.

To get physical quantities under interest, all above equations should be
numerically solved in a self-consistent manner.

\section{Results and Discussions}

To proceed the numerical calculations, we need to make some assumptions.
Since the number of the above self-consistent equations nonlinearly
increases with increasing the number of wave vectors of electrons and the
number of spin-wave modes in the middle FM region, which makes the
calculations too complicated to perform, for the sake of simplicity but
without losing the generality, we shall only consider the situations where
both the numbers of $k$ and $q$ taken in the following calculations are not
so large that the numerical calculations can be readily proceeded. This is
plausible, because the magnon-assisted transport property mainly depends on
the low-lying quantum well states of electrons in the middle FM, and only
the lower modes of spin waves are easy to emit\cite{Berger}, leading to
small energy levels of magnons\cite{Zeng}. Besides, considering that only
those electrons near the Fermi surface participate in the tunneling process,
we may take $\varepsilon _{k\sigma }\approx \varepsilon _{k_{F}}-\sigma M$,
denoted by $\varepsilon _{\uparrow }$ and $\varepsilon _{\downarrow }$ for
spin up and down electrons, respectively. In addition, we suppose that the
two side FM electrodes are made of the same materials, i.e., $M_{L}=M_{R}$, $%
P_{L}=P_{R}=P$, where $P_{L(R)}=[\Gamma _{L(R)\uparrow }-\Gamma
_{L(R)\downarrow }]/[\Gamma _{L(R)\uparrow }+\Gamma _{L(R)\downarrow }]$ is
the polarization of the left (right) FM layer. Then, the linewidth function
can be written as $\Gamma _{L\uparrow ,\downarrow }=\Gamma _{R\uparrow
,\downarrow }=\Gamma _{0}(1\pm P)$, where $\Gamma _{0}=\Gamma _{L(R)\uparrow
}(P=0)=\Gamma _{L(R)\downarrow }(P=0)$ will be taken as an energy scale. In
the following, we will take $P=0.7$, $k_{B}T=0.04\Gamma _{0}$, $I_{0}=\frac{%
e\Gamma _{0}}{\hbar }$ and $G_{0}=\frac{e^{2}}{\hbar }$ will be taken as
scales for the tunnel current and the differential conductance, respectively.

\subsection{Effect of Magnon-Assisted Tunneling}

In order to study whether the quantum oscillations of the conductance and
TMR observed in the FM-FM-FM tunnel junction are induced by spin-wave
excitations owing to spin accumulation, let us first examine the bias
dependence of the transport properties by considering the effect of
magnon-assisted tunneling in the AP state.

For given energy levels of the electrons and magnons in the middle FM
region, the bias dependent tunnel current ($I/I_{0}$), the differential
conductance ($G/G_{0}$) and $TMR$ for different $\gamma $ $(=T^{J}/T^{d})$
are shown in Fig. 1. It is observed that the tunnel current increases
nonliniearly with increasing the bias voltage. The twisted behavior of $%
I/I_{0}$ for $\gamma \neq 0$, presented in Fig. 1(a), comes from the
magnon-assisted tunneling, as no such behaviors are found for $\gamma =0$ in
either AP or P state. This can be clearly seen from the bias dependence of
the differential conductance, shown in Fig. 1(b), where the peaks and dips
appear for appropriate $\gamma $. Correspondingly, the TMR shows oscillating
behavior with increasing the bias voltage, as demonstrated in Fig. 1(c).
This observation manifests that the quantum oscillations of the conductance
and TMR in the FM-FM-FM tunnel junction can be caused by spin-wave
excitations, because when we turn off the effect of spin-wave excitations,
the oscillating behaviors of $G$ and $TMR$ disappear. Note that the peak in $%
G/G_{0}$ and one dip and one peak in $TMR$ are from the quantum resonant
tunneling of electrons. A larger TMR can be obtained for large $\gamma $,
and due to the s-d exchange interactions that could lead to spin-flip
scatterings, the TMR can be negative, as presented in Fig. 1(c).

\begin{figure}[tbp]
\vspace{-1.0cm} \includegraphics[width=0.6\linewidth,clip]{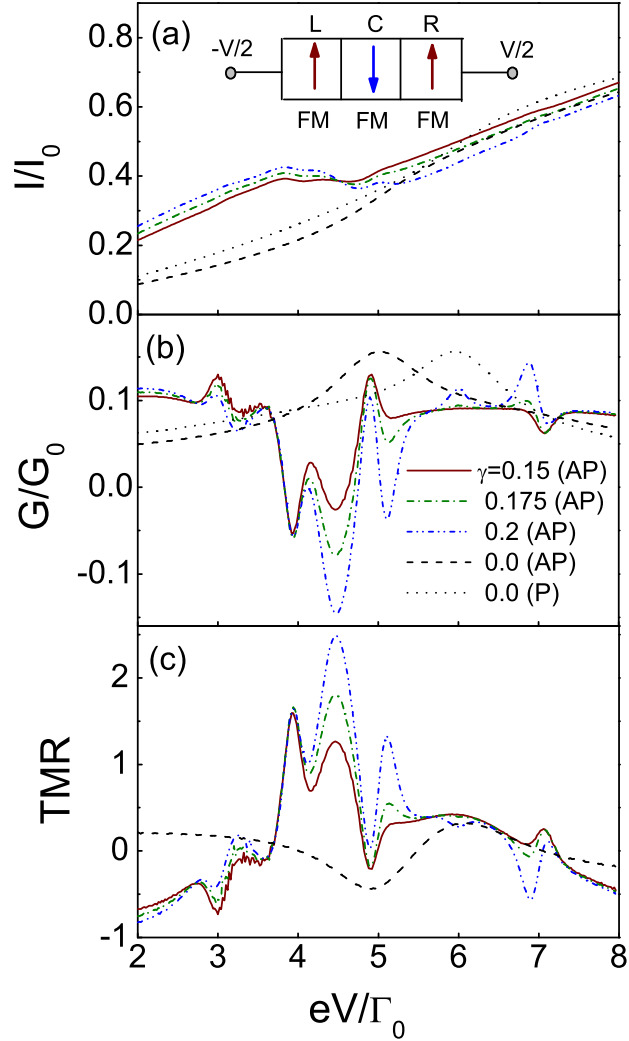} \vspace{%
-1.0cm}
\caption{(Color Online) Bias dependence of (a) the tunnel current I, (b) the
differential conductance G, and (c) TMR for the different $\protect\gamma $,
where $\hbar \protect\omega _{q}=0.5\Gamma _{0}$, $\protect\varepsilon %
_{\uparrow }=3.0\Gamma _{0}$ and $\protect\varepsilon _{\downarrow
}=2.0\Gamma _{0}$. }
\end{figure}

It should be remarked that the above oscillating behaviors of $G$ and $TMR$
appear only when $\gamma $ is in a suitable range, say, when $\gamma $ is
too small, no oscillations can be observed, while $\gamma $ is too large,
the self-consistent equations have no solutions.

The reason for the appearance of oscillations is that, when the applied bias
voltage exceeds a certain value, the non-equilibrium spin density can be
accumulated in the middle FM region, and spin waves can be excited. When
polarized electrons from the left FM layer tunnel into the central FM layer,
they are subject to scatterings from not only the polarized electrons in the
central region but also the spin waves from the accumulation owing to s-d
exchange interactions. It is possible that the electrons may turn their spin
directions by emission and absorption of magnons, leading to that the tunnel
conductance and TMR oscillate under a combination of effects of
magnon-assisted tunneling as well as quantum resonant tunneling\cite{mu2}
through the quantum well states in the central FM region.

Apparently, the polarized electrical current can excite magnons, while these
magnons participate in the tunneling process and in turn influence the
tunnel current. The number of magnons must be estimated self-consistently.
As an example, in Fig. 2, the bias dependent of the number of magnons $N$
for some parameters is presented. We may see that the number of magnons
oscillates with the bias voltage, which could be the main reason for the
oscillatory transport property of the system.

\begin{figure}[tbp]
\vspace{-1.0cm} \includegraphics[width=0.6\linewidth,clip]{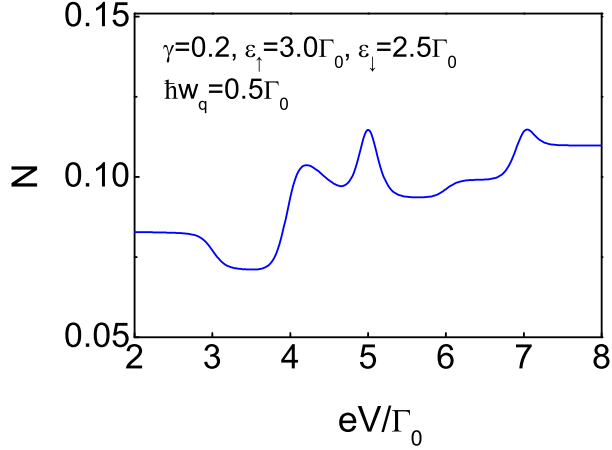} \vspace{%
-1.0cm}
\caption{(Color Online) Bias dependence of the number of magnons N in the AP
state. }
\end{figure}

\subsection{Effect of Magnon Modes}

There are a number of factors including magnon energy $\hbar \omega _{q}$
that can affect the transport behavior of the FM-FM-FM tunnel junction. The
bias dependence of the current, the different conductance and the TMR for
different magnon energies is shown in Fig. 3. It can be found that with
increasing $\hbar \omega _{q}$, apart from some quantitative changes of peak
positions and amplitudes, there are not much qualitative changes of the
current, conductance and TMR. Therefore, for a given magnon mode, the magnon
energy does not affect qualitatively the transport oscillating behavior of
the system. It is noted that, while some positions of the peaks and dips of $%
G$ and $TMR$ are influenced by the magnon energy, the others are not. This
observation indicates that the magnon energy is only one of factors
determining the positions and amplitudes of the peaks.

\begin{figure}[tbp]
\vspace{-1.0cm} \includegraphics[width=0.6\linewidth,clip]{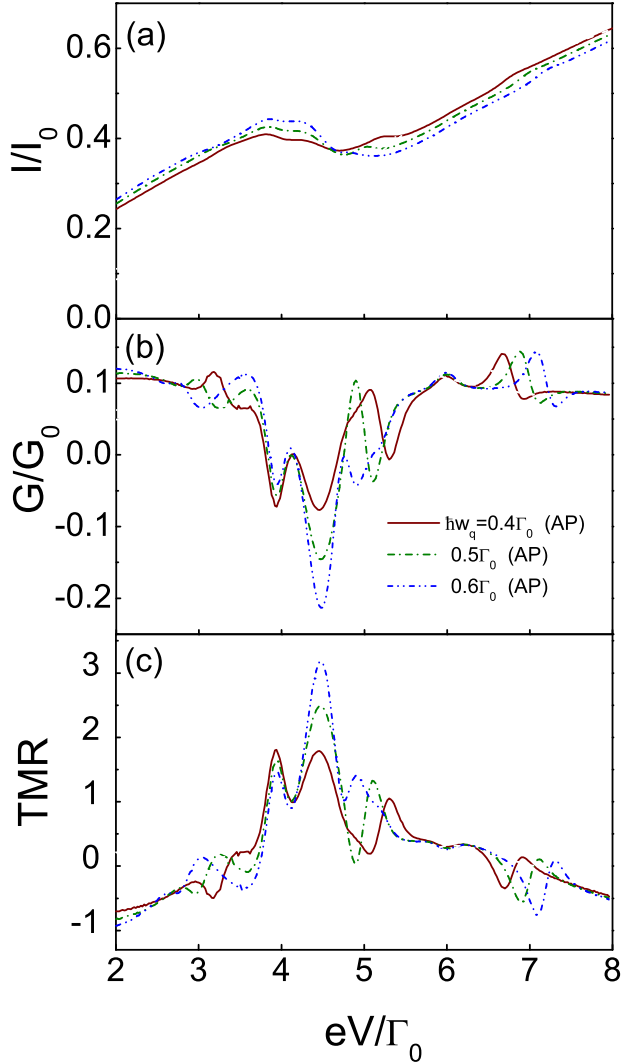} \vspace{%
-1.0cm}
\caption{(Color Online) Bias dependence of (a) the tunnel current I, (b) the
differential conductance G and (c) TMR for the different $\hbar \protect%
\omega _{q}$, where $\protect\gamma =0.2$, $\protect\varepsilon _{\uparrow
}=3.0\Gamma _{0}$ and $\protect\varepsilon _{\downarrow }=2.0\Gamma _{0}$.}
\end{figure}

For the sake of simplicity, in the aforementioned analysis we have adopted a
single spin-wave mode. Whether are the transport properties of the system
much affected qualitatively when we take more spin-wave modes into account?
The answer is presented in Fig. 4, where we have taken two and three
spin-wave modes to get the tunnel current, differential conductance and TMR.
For a comparison, we have also included the case with single mode. One may
see that at low biases, the magnon modes do not have so much effect on the
behaviors of $I/I_{0}$, $G/G_{0}$ and $TMR$, but at higher voltages the
magnitudes of the current, differential conductance as well as TMR change
somewhat remarkably. This is because at low biases the spin accumulation
effect is small, and the interaction between tunneling electrons and
spin-wave modes is weak, leading to the transport properties less
influenced; at large biases the spin accumulation effect becomes more
pronounced, and the interactions between electrons and magnons are strong,
the transport behaviors of the system are thus altered quantitatively. Note
that the round peaks and dips in the curves of the tunnel current shown in
Fig. 4 may come from combinations of the magnon-assisted tunneling as well
as the quantum resonant tunneling.

\begin{figure}[tbp]
\vspace{-1.0cm} \includegraphics[width=0.6\linewidth,clip]{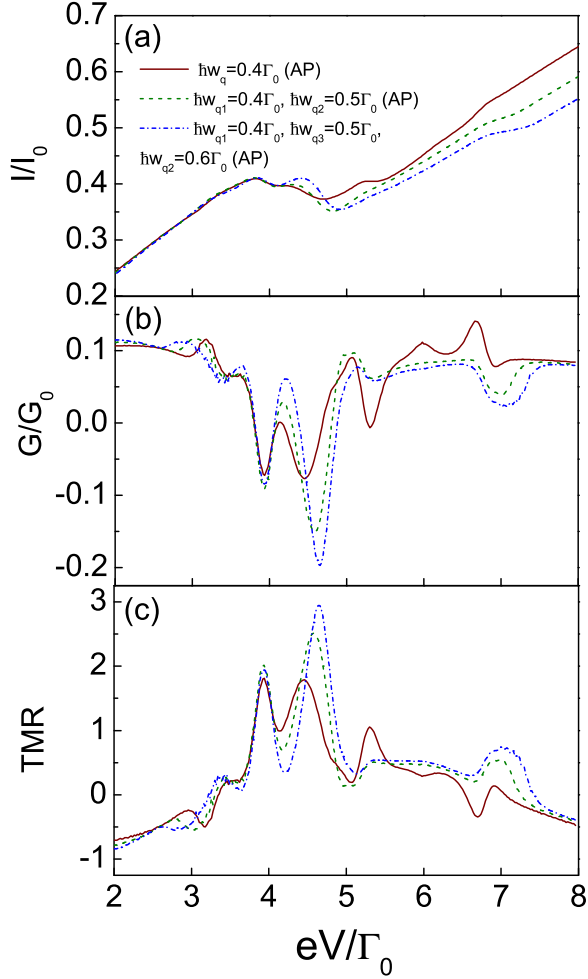} \vspace{%
-1.0cm}
\caption{(Color Online) Bias dependence of (a) the tunnel current I, (b) the
differential conductance G, and (c) TMR for different spin-wave modes, where
$\protect\gamma =0.2$, $\protect\varepsilon _{\uparrow }=3.0\Gamma _{0}$ and
$\protect\varepsilon _{\downarrow }=2.0\Gamma _{0}$. }
\end{figure}

\subsection{Effect of Electron Level in the Middle FM Layer}

The bias dependence of the current, the differential conductance and the TMR
for different energy levels ($\varepsilon _{_{k_{F}}}\equiv \varepsilon $)
of the electrons in the middle FM region is given in Fig. 5. It can be
observed that with lifting the energy levels of electrons in the central FM
layer, the peak and dip positions of $I/I_{0}$, $G/G_{0}$ and $TMR$ change
dramatically with increasing the bias voltage, while the shapes of the
curves retain quite similar for different energy levels of electrons in the
central region. This signifies that the energy levels of electrons in the
middle FM region do not affect the oscillating behavior itself of the
transport properties, but affect the positions of oscillating peaks and
dips. This can be understandable, because the transport behavior of
electrons are mainly determined by the scatterings from polarized electrons
and magnons to that the electrons are subject, when the energy levels of
electrons in the middle region are promoted, the resonant energies in
magnon-assisted and quantum resonant tunneling processes become different,
resulting in the behaviors shown in Fig. 5.

\begin{figure}[tbp]
\vspace{-1.0cm} \includegraphics[width=0.6\linewidth,clip]{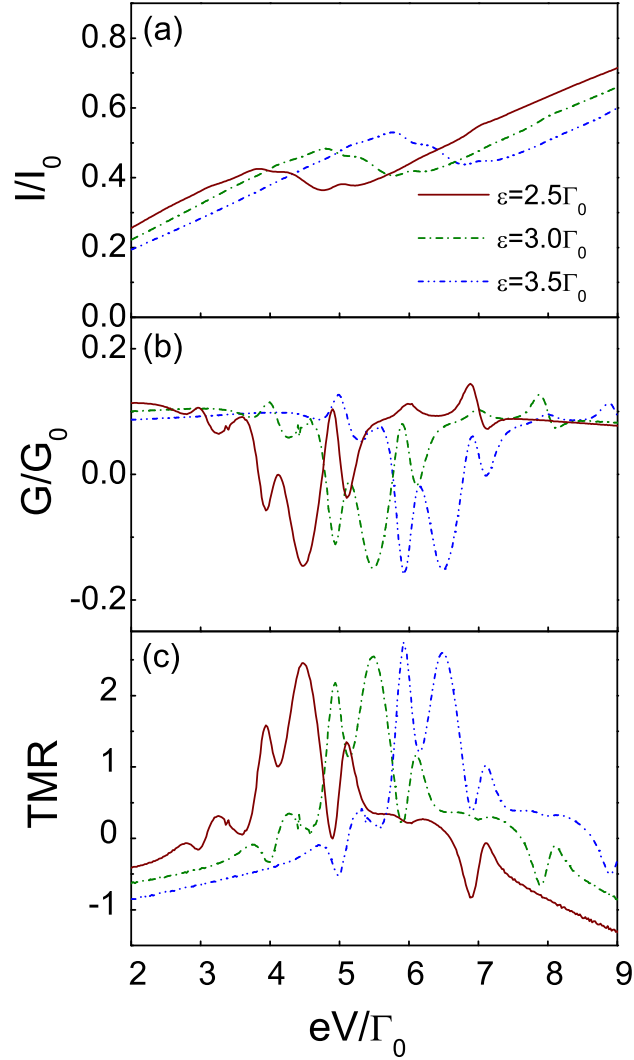} \vspace{%
-1.0cm}
\caption{(Color Online) Bias dependence of (a) the tunnel current I, (b) the
differential conductance G, and (c) TMR in the AP state for different energy
levels of electrons in the middle FM region, where $\protect\gamma =0.2$, $%
\hbar \protect\omega _{q}=0.5\Gamma _{0}$ and $M=1.0\Gamma _{0}$.}
\end{figure}

\subsection{Effect of Molecular Field in the Middle FM Layer}

The molecular fields of the middle FM layer could also have effect on the
transport properties of the system. The bias dependence of the current,
differential conductance and TMR for different molecular fields of the
central FM region is shown in Fig. 6. It is seen that as the molecular field
increases, the magnitude of the tunnel current becomes smaller, and the
oscillations of the differential conductance as well as TMR become more
apparent, where not only the number but also the positions of the
oscillating peaks change with increasing the molecular fields. This fact
suggests that the level spacing between the majorty and minority subbands of
electrons in the central region plays an important role in the oscillating
behaviors of the differential conductance and TMR in the present system.

\begin{figure}[tbp]
\vspace{-1.0cm} \includegraphics[width=0.6\linewidth,clip]{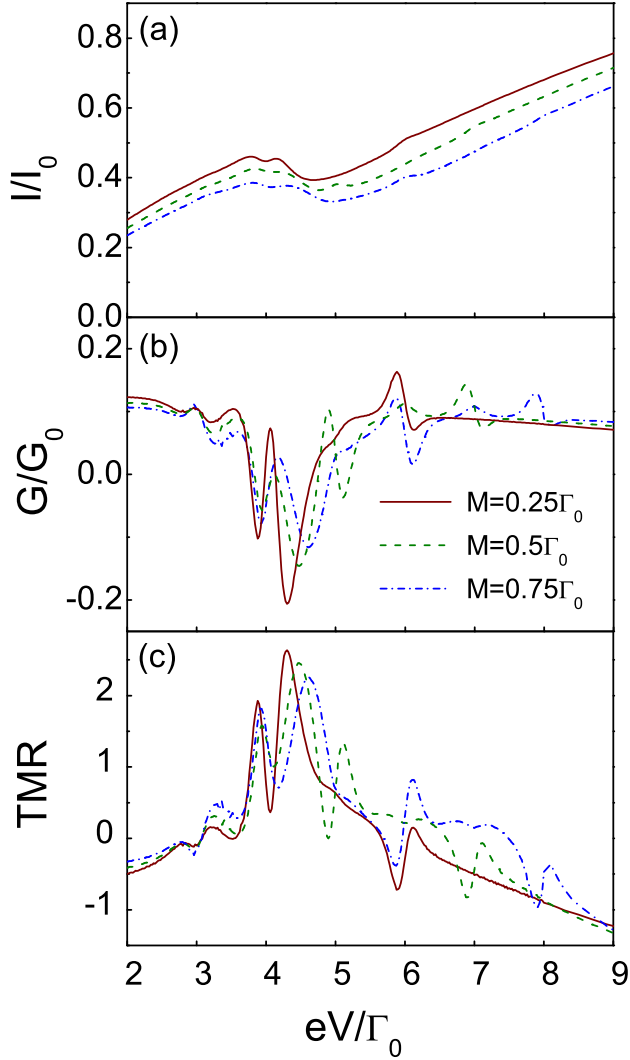} \vspace{%
-1.0cm}
\caption{(Color Online) Bias dependence of (a) the tunnel current I, (b) the
differential conductance G, and (c) TMR for different molecular fields of
the middle FM layer, where $\protect\gamma =0.2$, $\protect\varepsilon %
=2.5\Gamma _{0}$ and $\hbar \protect\omega _{q}=0.5\Gamma _{0}$.}
\end{figure}

\section{Summary}

In summary, we have probed the possibility of quantum oscillations of the
differential conductance and TMR in the FM-FM-FM tunnel junction in the AP
state. By self-consistently taking the s-d exchange interactions between
conduction electrons and the nonequilibrium spin density induced by spin
accumulation in the middle FM layer into account, we have found that the
differential conductance and TMR indeed oscillate with increasing the bias
voltage, thereby theoretically confirming qualitatively the inferrer and
experimental results presented in Ref. \cite{Zeng}. It has been unveiled
that the average number of magnons oscillates with the bias, which could be
the main reason for the oscillations of the conductance and TMR of the
system. When we turn off the s-d exchange interactions, i.e., taking $\gamma
=0$, no oscillations of the conductance and TMR were observed, showing that
the oscilations are indeed caused by the spin-wave excitations induced by
spin accumulations. In the P state, owing to the absence of spin accumulation%
\cite{jin3,jin4}, no oscillations of the conductance and TMR with the bias
can be found. We have also investigated the effects of the magnon modes, the
energy levels of electrons as well as the molecular field in the middle FM
region, and found that in spite of changes of the positions and amplitudes
of the oscillating peaks and dips, the oscillatory behavior of the transport
properties is not qualitatively affected. We anticipate that our findings
could offer clues for better understanding the experimental observation
presented in Ref. \cite{Zeng}.

Finally, we would like to remark that our preceding discussions could be
applicable to the system with a magnetic quantum dot coupled to two
ferromagnetic electrodes, where the oscillatory behavior of the transport
properties with the bias would be expected if the spin accumulation effect
is not neglected. The work toward this direction is under progress.

\acknowledgments

We are grateful to S. S. Gong, B. Gu, X. F. Han, H. F. Mu, Z. C. Wang and Q.
B. Yan for helpful discussions. This work is supported in part by the
National Science Fund for Distinguished Young Scholars of China (Grant No.
10625419), the National Science Foundation of China (Grant Nos. 90403036 and
20490210), and by the MOST of China (Grant No. 2006CB601102).

\end{document}